\documentclass[12pt]{article}
\usepackage[utf8]{inputenc}
\usepackage[english]{babel}
\usepackage{amsthm}
\usepackage{graphicx}

\usepackage{algorithm}
\usepackage{algpseudocode}
\usepackage{dsfont}
\usepackage{amssymb}
\usepackage{float}
\usepackage{xcolor}
\usepackage{amsmath}
\newtheorem*{remark}{Remark}
\DeclareMathOperator*{\argmax}{arg\,max}

\usepackage[round]{natbib}
\usepackage{amsmath}   

\begin{document}


\begin{center}
{\Large\bf Identification of mineralization in geochemistry along 
a transect based on the spatial curvature of log-ratios}
\end{center}
\begin{center}
{\sc Dominika Mikšová, Christopher Rieser, Peter Filzmoser }\\
{\it TU Wien\\
Vienna, Austria\\}
e-mail: {\tt dominika.miksova@tuwien.ac.at}
\end{center}

\begin{abstract}
Detecting subcropping mineralizations but also deeply buried mineralizations is one important goal in geochemical exploration. The identification of useful indicators for mineralization is a difficult task as mineralization might be influenced by many factors, such as location, investigated media, depth, etc. We propose a statistical method which indicates chemical elements related to mineralization along a transect. Moreover, the method determines along a transect the potential area of the deposit. The identification is based on General Additive Models (GAMs) for the element concentrations across the spatial coordinate(s). The log-ratios of the GAM fits are taken to compute the curvature, where high and narrow curvature is supposed to indicate the mineralization area. By defining a measure for the quantification of high curvature, the log-ratios can be ranked, and elements can be identified that are indicative of the anomaly patterns.

\end{abstract}

\section{Introduction}

Identifying geochemical processes as mineralization is defined as the presence of higher concentrations of particular chemical elements compared to the background concentration. However, it is challenging to define the background concentration,
since the threshold between background and mineralization will in general
not be characterized by a single number \citep[see][]{reimann2005geochemical}.
Nevertheless, it would be expected that a biogeochemical anomaly in mineralization exploration is indicated by a rapid spatial change in the concentration on top of the mineralization, depending on the type and extent of the mineralization. 
Nowadays, identifying geochemical features related to geochemical signature, 
and the separation of background and target zones of future local mineral 
exploration are becoming popular challenges in geochemistry.

Geochemical data in the form of chemical element concentrations are naturally compositional data, which are strictly non-negative values, forming parts of a whole. In this context we talk about compositional data analysis, and the 
log-ratio methodology introduced by \cite{Ait86} is the most common
approach in this context.
The important information to be analyzed is reflected in the log-ratios between the variables rather than in the absolute values. This ``relative'' information is
employed for a proper understanding of the data. 

There are two main difficulties with such an approach for practical geochemical
data sets: (a) Nowadays it is possible to measure the concentration of
dozens of chemical elements, and this leads to hundreds (or more) possible
pairwise log-ratios. Filtering out the elements which may indicate mineralization
is thus challenging. (b) Especially for mineral exploration there might not 
be many observations available, because often they are the result of a 
pre-study of the area. This creates further difficulties for the prediction 
of the location of a potential mineralization.

Since the identification of mineralization is a very relevant topic in practice, there are 
numerous publications available in the literature. This problem is also known under 
geochemical anomaly mapping, referring to a map presentation of geochemical uni-element or multi-element soil or plant data.
Related to the log-ratio methodology, the works of 
\cite{geochemCoDa}, \cite{carranza2017geochemical} and \cite{tolosana2014towards} aim to predict an anomalous presence of a mineral commodity. 
In these papers, a log-ratio transformation is applied, and then different mapping 
techniques are used to reveal the mineralization. The first mentioned paper uses centered log-ratio (clr) coefficients and isometric log-ratio (ilr) coordinates, since these representations
preserve metric properties. Anomalous compositions then originate from the robust barycentre. Using robust methods, the variables are split into two groups, then the variation of log-ratios gives ratios being in geochemical relationships, however the interpretation can be relatively weak and not specific enough. The work \cite{carranza2017geochemical} uses enrichment factors and log-ratios, where log-ratios are in terms of ilr coordinates. Based on kriging, a spatial correlation analysis was performed, where ilr values have much stronger positive spatial correlation with the known gold deposits. The paper also concludes that for mapping of significant anomalies, it is better to use ilr-transformed soil geochemical data than enrichment factors. A limitation is that this procedure
is a supervised method, meaning that the deposits need to be known for the input. 
Another example, proposed in \cite{tolosana2014towards}, shows that compositional data analysis 
is useful as a first step to identify geochemical features linked to natural phenomena. Logistic regression-like techniques are used to obtain a combination of variables that favor the presence of mineralization. Geostatistics is used to interpolate the composition to unsampled locations. 
However, the two proposed methods -- the Fisher method and Poisson processes --
can lead to incomparable results. The methods rather rely on information about high log-ratios, not in consideration of any spatial changes. However, this still gives informative results combining potential areas of interest and also relevant favorability in sense of ratios. 

The idea behind the presented method is that pairwise log-ratios would 
rapidly change towards a mineralized area. A rapid change would imply that the 
values of the log-ratio show strong curvature.
However, based on the observation data, ``curvature'' can only be computed
numerically, and this is infeasible if the rapid change is expressed only by
very few observations. For this reason we will approximate the underlying 
element concentration data by smooth values, producing a continuous signal
which allows to extract as many data points as necessary to compute a curvature
later on. Smoothness is important at this stage, because otherwise 
one could obtain arbitrary jumps in the log-ratios, leading to artifacts 
in the curvature. One could also argue that the smoothed concentration values
allow to suppress the effect of measurement uncertainties.

We decided to use Generalized Additive Models (GAMs) for the estimation of a 
smooth signal \citep{woodGam,yee2015vector}. The smoothness can be regulated by
a tuning parameter, which is selected by cross-validation. A GAM fit is based
on natural cubic splines with knots at every data point. Depending on the tuning
parameter, one can obtain the whole spectrum from the linear fit to the 
very non-smooth exact fit.
Once the GAM fits for both input variables of the log-ratio are available, 
the curvature of the log-ratio of the smoothed concentrations can be computed,
and an unsupervised learning method is employed, leading to a hitlist of 
log-ratios most suitable for finding mineralization. 
The proposed method has been tested on two real data sets, where the mineralizations 
are known, and the results seem to be reliable and carrying out promising prospects.

The paper is organized as follows: Section 2 introduces the methodology and provides 
a closer description of GAMs, additional information on the concepts of the curvature 
and its measurement. A detailed algorithm for the whole procedure of ranking of 
log-ratios is proposed in Section 3. Section 4 demonstrates numerical 
experiments based on two real geochemical data sets. The last Section 5 concludes 
and provides possible extensions of the proposed method.

\section{Methodology}\label{sec:method}

\subsection{Motivating artificial example}

The principal idea is to investigate the curvature of log-ratios. The higher an index involving
the curvature, the more likely a point of mineralization has been identified.

\begin{figure}[h] 
\centering
\includegraphics[width=1\textwidth]{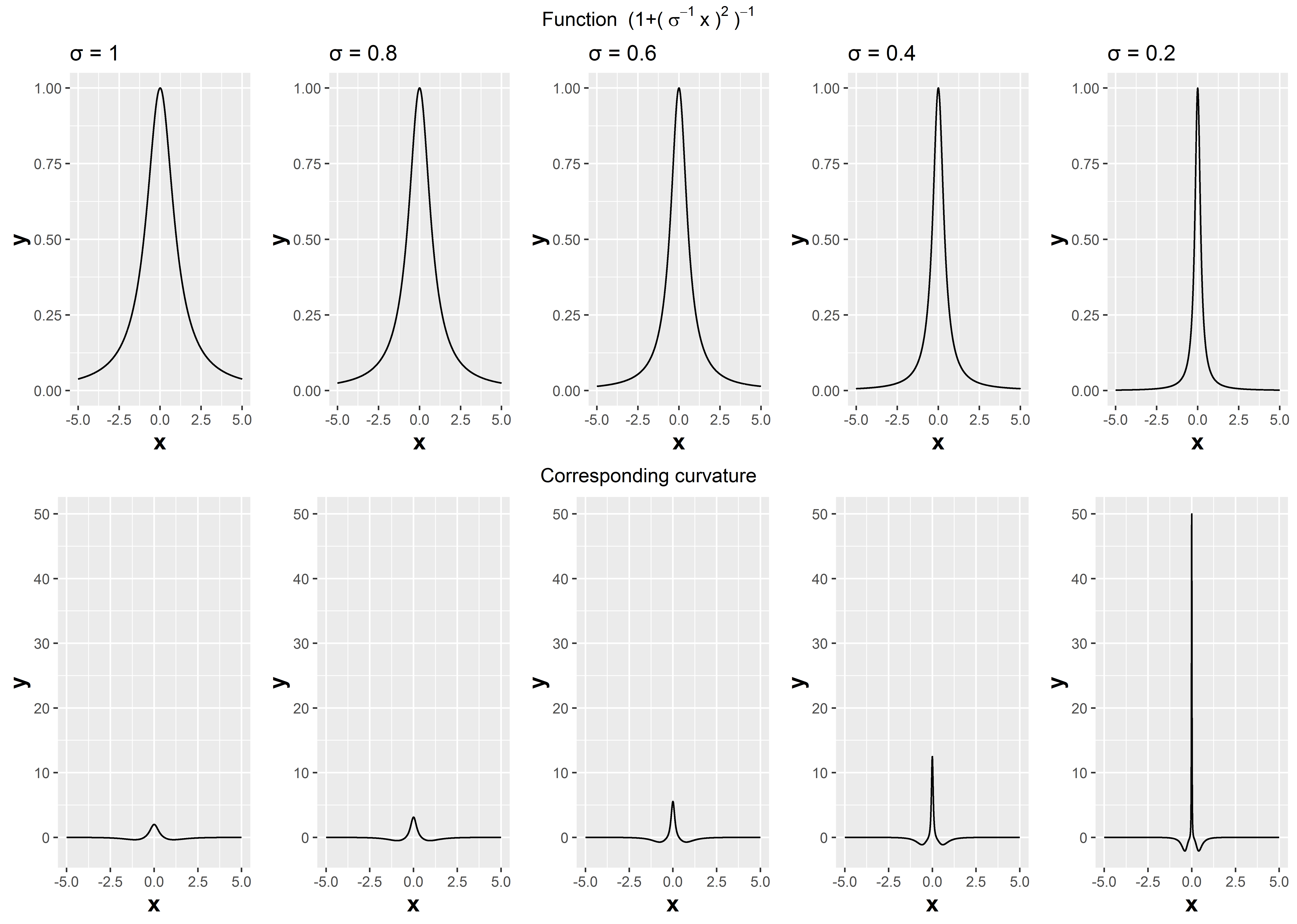}
\caption{Top row: function $x \mapsto (1+ (\frac{x}{\sigma})^2)^{-1} $ for different $\sigma$. Bottom row: corresponding curvature (to be defined in Section 2.3).}
\label{fig:illustrationCurvature}
\end{figure}

For illustration purposes, let us consider the function $x \mapsto (1+ (\frac{x}{\sigma})^2)^{-1} $, 
for fixed $\sigma > 0$. Figure \ref{fig:illustrationCurvature} shows on the top row the function itself 
for different values of $\sigma$, and the bottom row presents the corresponding curvatures. 
An appropriate measure of curvature will be defined later in Section 2.3.
As it can be seen on the top row, the lower $\sigma$ becomes, the quicker the spatial change of the function is. 
Looking at the bottom row, this translates into a growing value of the curvature at zero. 
The case of $\sigma \to \infty$, not displayed here, corresponds to the function being constant equal 
to one, thus having zero curvature. The methodology developed in the following involves the 
curvature as a measure of how quick a signal undergoes spatial change. Of course, this measure needs to be normalized appropriately
in case of several peaks with possibly different curvature.

\subsection{GAMs}

Since in many studies only few observations are available, 
the original concentration data are approximated by smooth curves originating from
Generalized Additive Models (GAMs), before log-ratios for computing curvature are considered.
This is preferable to computing a curvature measure directly from the log-ratios
of the observations, since with smooth fits one can in principle generate arbitrarily many
observations, and derive a more stable value for the curvature.
GAMs have the advantage that the degree of smoothness of the fit to the data can be
tuned.

Denote $(x_1,y_1),\ldots ,(x_n,y_n)$ the $n$ observed data of the measured concentration $y$ at position $x$ 
of a certain element, where we assume that $x_1 \leq \dots \leq x_n$ are measured along a 
linear transect. At the heart of GAMs is a weighted penalized log-likelihood problem over a suitable function space $\mathcal{H}$, estimating the presumably smooth linear predictor $\eta$
\begin{align} \label{splines}
\hat{\eta} = \argmax_{\eta \in \mathcal{H}} \, \sum^{n}_{i=1} \omega_i l(y_i|x_i;\eta) - \lambda \int (\eta''(x))^2 dx,
\end{align}
where  $\lambda$ is the so called smoothing parameter, $l$ is the log-likelihood function, and $\omega_i$ are predefined weights.
To explain this further, we note that in the GAM framework one necessary assumption is that the response $y$ belongs to the exponential family, thus its density is of the form 
\begin{align*}
    f(y|x) = \exp \Bigg(\frac{ \theta y|x  - b(\theta)}{a(\psi)} + c(y|x,\psi)\Bigg) ,
\end{align*}
with parameters $\theta$, $\psi$ and given functions $a(\cdot)$, $b(\cdot)$ and $c(\cdot)$. It can be shown that with this assumption we are able to rewrite the log-likelihood in dependence of the conditional mean $\mathbb{E}(y|x)$. Given now a so called link function $h(\cdot)$ -- a smooth monotonic function which the user must normally choose, except in some special cases for which we get a canonical link -- we model the composition of the expectation of $y$ with $h(\cdot)$ as $h(\mathbb{E}(y|x)) = \eta(x)$ and are therefore able to implicitly write the log-likelihood in dependence of the linear predictor $\eta$. Modeling the mean in such a way, we are able, once $\eta$ is estimated, to make predictions of $y|x$ through $h^{-1}(\eta(x))$. The choice of $h(\cdot)$ is in many cases not crucial as long as its domain matches with the range of possible values of $\mathbb{E}(y|x)$.

The smoothing parameter $\lambda$ controls the trade-off between smoothness and the fit to the data; the bigger $\lambda$ becomes, the smoother the function will be, as in the case $\lambda \to \infty$ we get $\eta'' \equiv 0$ and therefore $\eta$ is a linear function. Typically, the smoothing parameter $\lambda$ is chosen in a data dependent way by using either Generalized Cross Validation (GCV) or Restricted Maximum Likelihood (REML).

For fixed $\lambda$, the function $\eta$ solving problem (\ref{splines}) can be written in terms of a cubic B-spline basis, see \cite{friedman2001elements}, for example. Therefore, $\eta(x) = \sum^n_{j=1} h_{j}(x)\beta_j$, where $h_{j}$ are the cubic B-spline basis functions, and $\beta_j$ the 
coefficients.

For a more thorough introduction to GAMs, GCV, REML and how problem (\ref{splines}) is solved algorithmically, we refer to \cite{woodGam}. \\

In the applications of our method to the data of Section 4, we decided to model $y_i|x_i$ belonging to the family of Tweedie distributions with a log-link function $h\equiv \log$. This means that we model 
\begin{align}
&h(\mathbb{E}(y|x))= \eta(x) \qquad \mathbb{V}(y|x)=\mathbb{E}(y|x)^p \frac{\psi}{\omega(x)} \label{mean-var-relation}
\end{align}
for some $p \in (1,2)$. Furthermore, to capture the effect of outliers in the response we use predetermined weights $\omega_i$;
for example we  took $\bar{\omega}_i:= \max\Big(\frac{|y_i-\hat{\mu}|}{\hat{\sigma}},1\Big)$, where $\hat{\mu}$ and $\hat{\sigma}$ are the sample mean and the sample standard deviation of $y$, and then put $\omega_i = \frac{\bar{\omega}_i}{\sum \bar{\omega}_i} $.\\
In these particular examples, the choice of this family of distributions and this specific link function is motivated by the fact that it provides a very flexible range of modelling the mean-variance relationship -- as it comprises many distributions through the additional parameter $p$. Furthermore, the predetermined weights have been chosen in a way such that outliers get upweighted, i.e. if a point $y_i $ is bigger than   $\hat{\mu} + \hat{\sigma} $ then it will get upweighted proportionally. This seems necessary as the variance will likely be higher on top of mineralizations and thus over- or under-dispersion in our model might otherwise appear. All in all, inspecting the linear predictor vs. the residuals as well as the fitted values vs. the response plots show consistency with our choice of link-function and thus also with model (\ref{mean-var-relation}). Some of these plots are shown in Section~4.


\subsection{Curvature of log-ratios}

Since we are interested in the log-ratios of two chemical elements, we
denote for an element $el_1$ and an element $el_2$ their respective GAM fits on the response scale 
$\hat{f}_{el_1}(x) :=h^{-1}(\hat{\eta}_{el_1}(x))$ and  $\hat{f}_{el_2}(x):= h^{-1}(\hat{\eta}_{el_2}(x))$.
The log-ratio of the fits
\begin{align*}
 g(x):=\log\bigg(\frac{\hat{f}_{el_1}(x)}{\hat{f}_{el_2}(x)}\bigg)= \log\big(\hat{f}_{el_1}(x)\big) - \log\big(\hat{f}_{el_2}(x)\big)
\end{align*}
is then shifted and scaled to  $ c  (g(x)-\min g(x)) $,
with the scaling constant \\$c:=|\max_{x \in [x_1,x_n]}g(x)-\min_{x\in [x_1,x_n]}g(x)|^{-1}$, whenever $g$ is not constant. 
This is done to make our method comparable across different log-ratios. In the special case of $g$ being constant we can set $c$ to one, as such functions will be ranked lowest by the measure described below. 

As a next step we will define a measure for identifying important log-ratios based on the curvature. The curvature $\kappa$ of the shifted and scaled function is defined as  
\begin{align}
        \kappa(x):= \frac{|cg''(x)|}{(1+(cg'(x))^2)^{\frac{3}{2}}} \label{defKappa},
\end{align}
see \cite{Curvature},
and thus, as
\begin{align*}
& g'(x) = \frac{\hat{f}'_{el_1}}{\hat{f}_{el_1}}(x) - \frac{\hat{f}'_{el_2}}{\hat{f}_{el_2}}(x)\\
& g''(x)= \frac{\hat{f}''_{el_1}}{\hat{f}_{el_1}}(x) - \bigg(\frac{\hat{f}'_{el_1}}{\hat{f}_{el_1}}(x)\bigg)^2 - \frac{\hat{f}''_{el_2}}{\hat{f}_{el_2}}(x) + \bigg(\frac{\hat{f}'_{el_2}}{\hat{f}_{el_2}}(x)\bigg)^2
\end{align*}
holds, this amounts to calculating the first and second derivatives of the GAM fits on the response scale of the individual elements. 


\subsection{Curvature exceeding a threshold}

\begin{figure}[h] 
\centering
\includegraphics[width=1\textwidth]{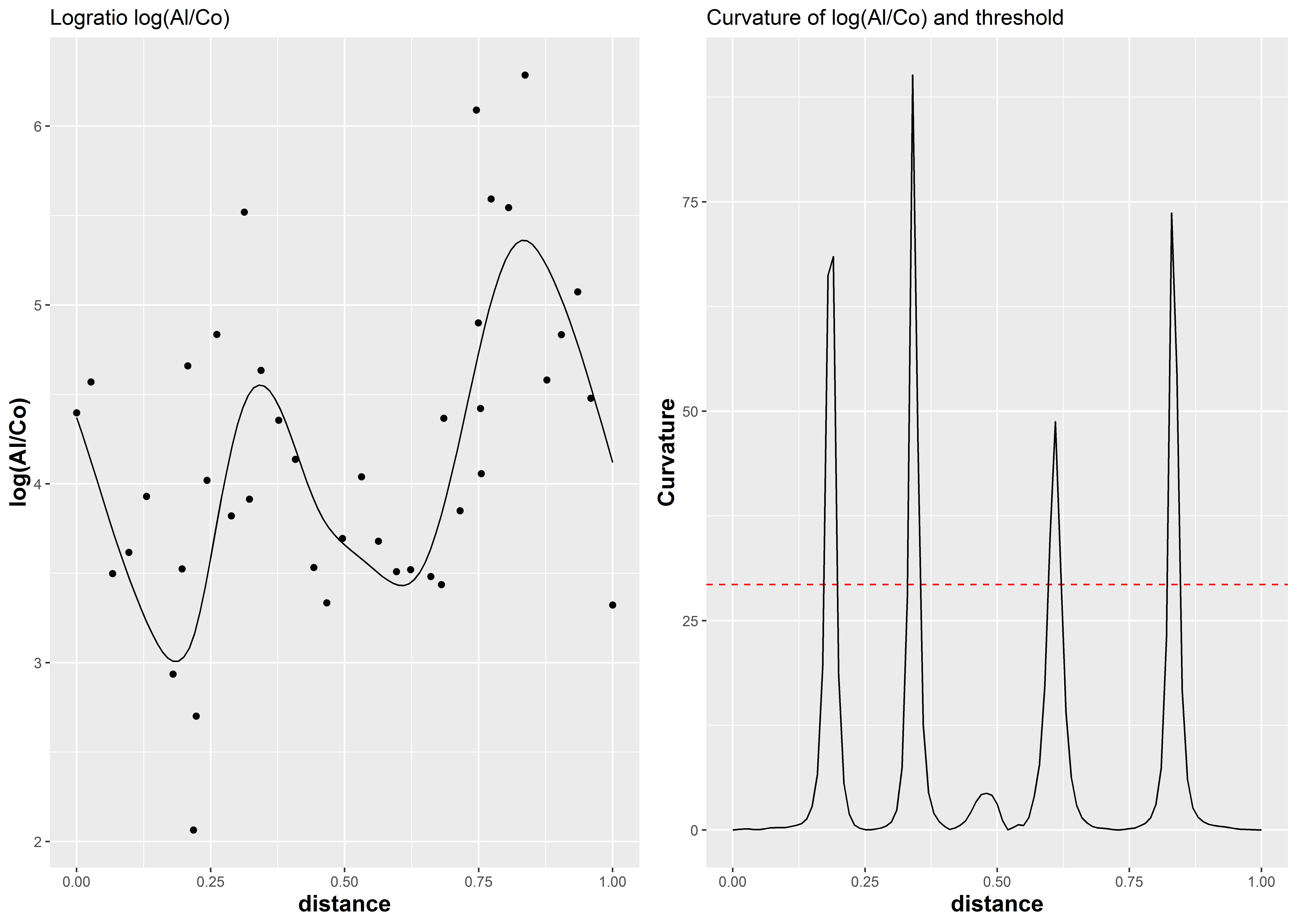}
\caption{Example of a log-ratio plot of the elements Al and Co and the corresponding curvature plot with the threshold (dashed line). One can see the correspondence between local maxima above the threshold in the right plot with the peaks in the left plot.}
\label{fig:thresholdCurvature}
\end{figure}

For each such a combination of elements we calculate the mean and the variance of the curvature $\kappa$, namely 
\begin{align}
    &\mu =  \int^{x_n}_{x_1} \kappa(x) dx   \label{meanCurv}\\ 
    &\sigma^2 = \int^{x_n}_{x_1} (\kappa(x)-\mu)^2 dx \label{sdCurv}
\end{align}
and define the set of crossings with the threshold $\mathcal{T}:= \mu+\sigma$ by
\begin{align} 
    \mathcal{C}:= \{ x \in[x_1,x_n] | \kappa(x)=\mathcal{T}  \} \bigcup  \{ x \in \{x_1,x_n\} |  \kappa(x)\geq \mathcal{T} \}, \label{setC}
\end{align}
where the second set contains $x_1$ and/or $x_n$ depending if they are above the threshold or not. The purpose of defining this set is to detect the points where the curvature $\kappa$ crosses the threshold and subsequently exceeds it, so that we are left with only a few high local maxima -- see for example Figure \ref{fig:thresholdCurvature}. Of course these maxima depend on the definition of the threshold, and it seems reasonable to take the mean plus the standard deviation over the whole range, because, as it is implied by Chebyshev's inequality, the further we get from this threshold the less likely an observation is. 

\begin{remark}
If the set $\mathcal{C}$  is finite it follows that its cardinality is even by definition.
For example, if the second set in (\ref{setC}) is empty and the first is non-empty, then 
the set must have even cardinality as otherwise before or after the last crossing it must always remain above the threshold, thus $\kappa(x_1)>\mathcal{T}$ or $\kappa(x_n)>\mathcal{T}$.  \\
\end{remark}

Instead of analytically solving the equation $\kappa(x)=\mathcal{T}$ we decided to uniformly sample a high number of points $N$ in the interval $(x_1,x_n)$ - where we always add  $\{ x \in \{x_1,x_n\} |  \kappa(x)\geq \mathcal{T} \}$ to the set - and then check for each of these points $x_l$ if $\kappa(x_l)-\mathcal{T}$ has a change of sign. If so, we add this point to the set and we denote $\hat{\mathcal{C}}$ the set constructed in this way. We do the latter so that we can continue to work with a finite set which is of even cardinality, and one could see the measure constructed 
in the following as an approximation to the analytical case.

\subsection{Measure for comparing curvature}

We define the measure to compare the curvature values of the log-ratio of
two elements $el_1$ and $el_1$ as
\begin{align}
    c(el_1,el_2):= \frac{2}{L} \sum^{\frac{L}{2}}_{l=1}  \max_{x \in [x_{j_{2l-1}},x_{j_{2l}}]} (\kappa(x)-\mathcal{T})_{+}^2, \label{measure}
\end{align}
where $(\cdot)_+$ denotes $\max(\cdot,0)$, and where $x_{j_1}<...<x_{j_l}$, $l=1,...,L$ are the ordered points of the finite set $\hat{\mathcal{C}}$.
We will call this measure $c$-value in the following.

The more the curvature $\kappa$ exceeds the threshold in $[x_{j_{2l-1}},x_{j_{2l}}]$, 
the bigger the maximum and thus also the measure will be. Therefore, a relatively fast change in the original signal will contribute a lot to this measure. 
By including the factor $\frac{1}{2 L}$, the measure $c(el_1,el_2)$ becomes the mean, 
and thus $c(el_1,el_2)$ is high if the peaks above the threshold $\mathcal{T}$ are high on average.

Since this measure is normalized, it can be used to compare all different pairs of log-ratios, and
it can even be used to compare different data sets taken at the same 
locations, such as measurements from different sample materials or soil layers.
The log-ratio pairs can be ordered according to the value of the measure, and 
the pairs corresponding to the highest ranking will be most promising for the 
identification of mineralization.


\section{Algorithm}

In the following we will describe the algorithm using the methodology above, which takes as 
an input 
the element concentrations for $n$ observations, denoted as the vectors
$\mathbf{y}_{el_1},\ldots ,\mathbf{y}_{el_m}$ of length $n$, where 
$(y_{el_k})_i$ is the $i$-th observation of the measured concentration of the $k$-th element, and the location vectors $\mathbf{x}$ of length $n$, where $x_i$ is the $i$-th observed location. 
The output is a matrix $C$ with entries $c(el_r,el_s)$ for different elements $el_r$ and $el_s$.  \\

\noindent
\textbf{Step 1:} Before fitting the GAM model we scale the entries of $\mathbf{x}$
to the range $[0,1]$, and then we calculate the weights $(\omega_{el_k})_i$ for the 
element concentrations (here for the $k$-th element), see definition below 
model~(\ref{mean-var-relation}). \\

\noindent
\textbf{Step 2:}
As a next step we fit a GAM model to each element, meaning that for the measurements 
$(x_i,(y_{el_k})_i)$ we solve 
\begin{align*}
\max_{\eta_{el_k}} \, \sum^{n}_{i=1} (\omega_{el_k})_i~ l((y_{el_{k}})_i|x_i;\eta_{el_k}) - \lambda \int (\eta_{el_k}''(x))^2 dx.    
\end{align*}
For the applications presented in Chapter 4 we have decided to use the Tweedie family and the log-link function. The fitting is done with the help of the R package \texttt{mgcv} \citep{woodGam}, and the smoothing parameter $\lambda$ is tuned automatically by using the implemented REML criterion. \\

\noindent
\textbf{Step 3:}
Once all the elements have been fitted, thus once we have computed all $\hat{\eta}_{el}$, we compute for a high number of points $x \in [x_1,x_n]$, typically we used N = 3000, all the possible shifted and scaled log-ratios at these points. Therefore, denoting $\mathcal{X}$ the ordered set of these points $x$, we calculate for each possible pair of elements $el_1$ and $el_2$, $c:=|\max_{x \in\mathcal{X}}g(x)-\min_{x\in\mathcal{X}}g(x)|^{-1}$, where $g(x)= \log(\hat{f}_{el_1}(x)) - \log(\hat{f}_{el_2}(x))$; $\hat{f}_{el_1}(x) :=h^{-1}(\hat{\eta}_{el_1}(x))$ and  $\hat{f}_{el_2}(x):= h^{-1}(\hat{\eta}_{el_2}(x))$. If $g$ is constant we set $c = 1$. \\

\noindent
\textbf{Step 4:}
As a next step we calculate the curvature of these log-ratios. Thus, for each such pair of elements $el_r$ and $el_s$ and all $x \in \mathcal{X}$ we need to determine $g'(x)$ and $g''(x)$ first. This is done numerically. For a small $\epsilon $,  say $10^{-3}$, we compute the approximate derivatives for all elements
\begin{align*}
    & \hat{f}'_{el}(x) \approx \frac{1}{2\epsilon} (\hat{f}_{el}(x+\epsilon)-\hat{f}_{el}(x-\epsilon)) \\
     &\hat{f}''_{el}(x) \approx \frac{1}{\epsilon^2} (\hat{f}_{el}(x+\epsilon) - 2 \hat{f}_{el}(x) +\hat{f}_{el}(x-\epsilon)) 
\end{align*}
and as
\begin{align*}
& g'(x) = \frac{\hat{f}'_{el_1}}{\hat{f}_{el_1}}(x) - \frac{\hat{f}'_{el_2}}{\hat{f}_{el_2}}(x)\\
& g''(x)= \frac{\hat{f}''_{el_1}}{\hat{f}_{el_1}}(x) - \bigg(\frac{\hat{f}'_{el_1}}{\hat{f}_{el_1}}(x)\bigg)^2 - \frac{\hat{f}''_{el_2}}{\hat{f}_{el_2}}(x) + \bigg(\frac{\hat{f}'_{el_2}}{\hat{f}_{el_2}}(x)\bigg)^2
\end{align*}
holds, it is easy to compute  $ \kappa(x):= \frac{|cg''(x)|}{(1+(cg'(x))^2)^{\frac{3}{2}}} $ for each pair of elements and $x \in \mathcal{X}$.\\

\noindent
\textbf{Step 5:}
After this we compute an approximation to the treshold $\mathcal{\tau}$ by approximating (\ref{meanCurv}) and (\ref{sdCurv}). Thus we define
\begin{align*}
  \mathcal{\tau} :=  \frac{1}{|\mathcal{X}|} \sum_{x \in \mathcal{X}} \kappa(x) + \sqrt{ \frac{1}{|\mathcal{X}|} \sum_{x \in \mathcal{X}} \bigg(\kappa(x) -  \frac{1}{|\mathcal{X}|} \sum_{x \in \mathcal{X}} \kappa(x)\bigg)^2} .
\end{align*}

\noindent
\textbf{Step 6:}
Next we compute an approximation to the set $\mathcal{C}$.  We define the set $\hat{\mathcal{C}}$ as all the $x$ for which we have $\kappa(x) = \mathcal{\tau}$ or for which $\kappa(x)$ is smaller than $\mathcal{\tau}$ and then the next element in $\mathcal{X}$, say $\bar{x}$, we have $\kappa(\bar{x}) > \mathcal{T}$. Also we add $x_1$ and/or $x_n$ if $\kappa$ is bigger or equal than $\mathcal{\tau}$ there. This seems like a reasonable approximation as long as the cardinality of $\mathcal{X}$ is high enough. \\

\noindent
\textbf{Step 7:}
Finally, we compute for each pair of elements the measure  $c(el_r,el_s):= \frac{2}{L} \sum^{\frac{L}{2}}_{l=1}  \max_{x \in I_l} (\kappa(x)-\mathcal{T})_{+}^2$, where $I_l := [z_{j_{2l-1}},z_{j_{2l}}]$, $z_j$ is a point of the set $\hat{\mathcal{C}}$  and where $L$ is the cardinality of $\hat{\mathcal{C}}$. \\

In short, the steps above can be subsumed into the following algorithm:

\begin{algorithm}[H]
\caption{Log-ratio measures}
\label{algorithm}
\begin{algorithmic}[1]
\For{k = 1,...,m}
\State Calculate  weigths, i.e.~calculate $\hat{\mu}$ and $\hat{\sigma}$ of $\mathbf{y}_{el_k}$ 
and set $(\omega_{el_k})_i:= \max(\frac{(y_{el_k})_i-\hat{\mu}}{\hat{\sigma}},1)$ 
\State Solve $\max_{\eta_{el_k}} \, \sum^{n}_{i=1} (\omega_{el_k})_i ~l((y_{el_{k}})_i|x_i;\eta_{el_k}) - \lambda \int (\eta_{el_k}''(x))^2 dx$
\EndFor
\For{(r,s) in \{1,...,m\}}
\State Calculate curvature $\kappa$ for the shifted and scaled $\log\Big(\frac{f_{el_r}}{f_{el_s}}\Big)$
\State Compute $\mu$ and $\sigma^2$ as described in (\ref{meanCurv}) and (\ref{sdCurv})  and set  $\mathcal{T}:= \mu + \sigma$ 
\State Draw uniformly $N$ points from $(x_1,x_n)$ and add crossing points of $\kappa$ with $\mathcal{T}$  \qquad \hphantom{bla} to $\hat{\mathcal{C}}=\{ x \in \{x_1,x_n\} |  \kappa(x)\geq \mathcal{T} \}$ 
\State Set $c(el_r,el_s):= \frac{2}{L} \sum^{\frac{L}{2}}_{l=1}   \max_{x \in I_l} (\kappa(x)-\mathcal{T})_{+}^2$
\EndFor
\State Define matrix $C$ with entries $c(el_r,el_s)$
 \State \textbf{return}  Matrix $C$
\end{algorithmic}
\end{algorithm}
\hphantom{...}

Once these values $c(el_r,el_s)$ are obtained for all element combinations, 
we can compute a ranked list for the log-ratios from highest to lowest, or a heatmap based on the matrix $C$ -- where we scale the entries by the maximum entry first. As the measure $c(el_r,el_s)$ is comparable across materials, it is also possible to use these matrices to 
explore accumulated heatmaps for a comparison of different plant materials, see next section.

The whole algorithm as described above has been implemented in the software environment {\tt R} \citep{RDev} using mainly the {\tt gam()} function implemented in the package \textbf{mgcv} \citep{mgcv}. This software is available
from the authors upon request.


\section{Experimental results}


The proposed method has been tested on two real geochemical data sets with known mineralization.
Both data sets are sampled along a (more or less linear) transect. The known 
locations of mineralizations can be used to evaluate the proposed procedure.

\subsection{Juomasuo data}

The Juomasuo data set is described in detail in \cite{middleton2018biogeochemical}, and it
originates from the UltraLIM project, where biogeochemical samples 
have been taken in the years 2013 and 2014 in a subarctic region in northern Finland. Juomasuo, among all available sites, is the largest of the known Au deposits in the area. We take the data from 2013, where three
different sample plant materials have been collected (Crowberry, Bilberry and Labrador 
tea). The investigated tissue of the plant species is either twig/stem or leaf/needle, and  
they have been analyzed for the concentration of more than 40 chemical elements. 
Depending on the plant material, 27 to 30 samples are available, and we focus on 
the concentration values of 27 chemical elements with reasonable data quality.
 Moreover, plants showed strong positive apical anomaly patterns for Cobalt (Co), Iron (Fe), Thorium (Th), Uranium (U) and rare earth elements (REE), such as Cerium (Ce), Lanthanum (La) and Neodymium (Nd) in more than one species. 

\begin{figure}[H] 
\centering
\includegraphics[width=1\textwidth]{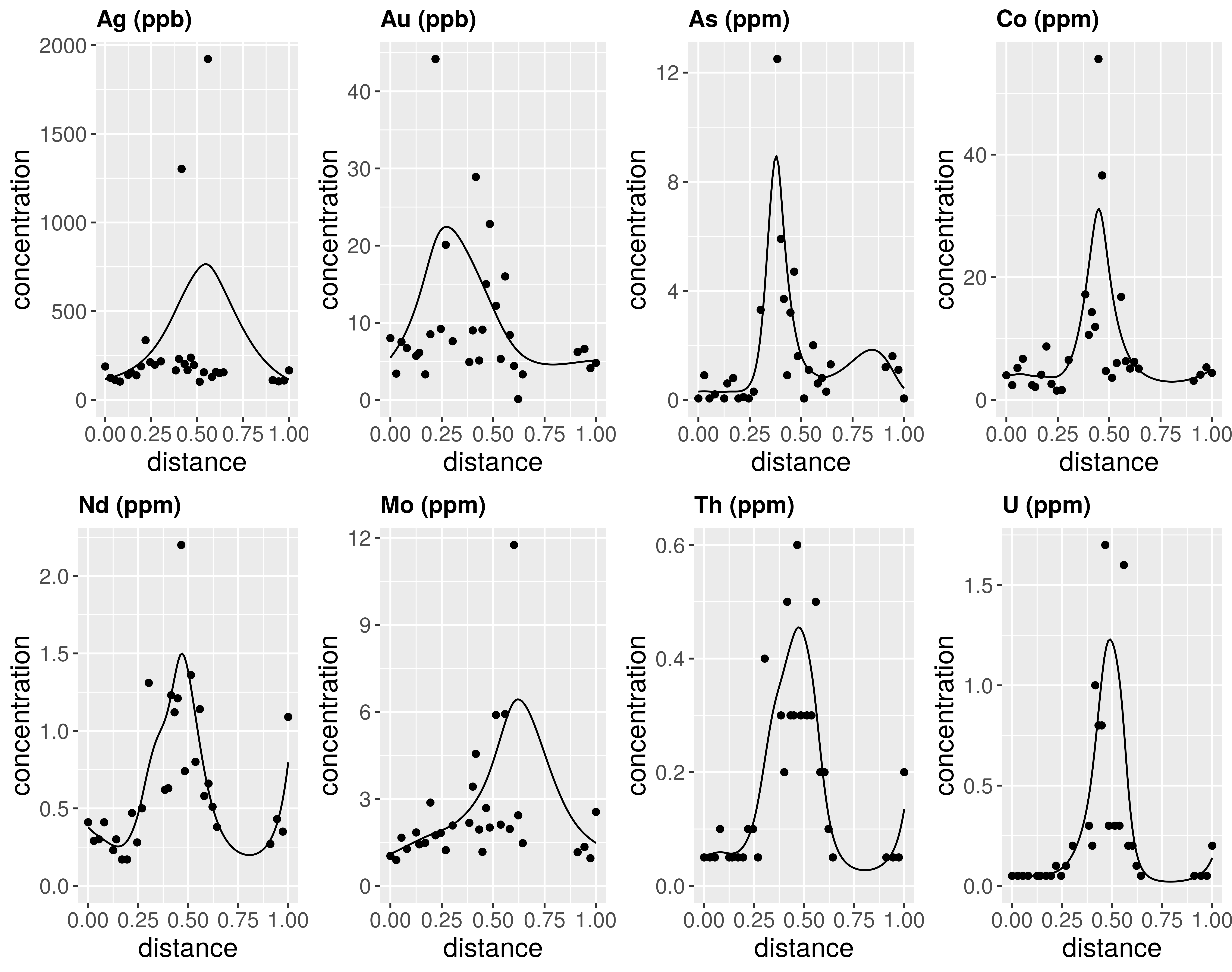}
\caption{GAM fits (lines) for eight selected elements measured in Crowberry twigs from the 
Juomasuo data set are displayed together with their original concentrations (dots).}  
\label{fig:gamfits_juom}
\end{figure}

The sample locations are approximately along a line, and as a first step the 
distances between the samples are computed.
The distance between the most extreme sites is 1271 meters, an on average one sample 
has been taken per 45 meters. For reasons of comparability, the distances are 
normalized from 0 to 1 as an input for the GAM models 
(see horizontal axis of Figure~\ref{fig:gamfits_juom}). 
The known mineralizations are at the following (normalized) distances:
0.3, 0.38, 0.43, 0.48, 0.51, 0.53 and 0.55.


Figure~\ref{fig:gamfits_juom} shows the GAM fits applied on eight selected variables
measured in Crowberry twigs,
together with the original concentration values (dots). 
The GAM fits result in a very smooth signal, even around normalized distance 0.75, where a gap in
the sampling procedure occurred due to a peat bog. As mentioned above, the known mineralizations
are at distances between 0.3 and 0.55, and this is visible also in Figure \ref{fig:gamfits_juom}, 
where one can clearly see anomaly patterns around these distances.
Due to the choice of the weights for the GAM fit, the outliers 
have a stronger impact on the fits, which is a desirable effect for the purpose of anomaly detection.

\begin{figure}[!ht] 
\centering
\includegraphics[width=1\textwidth]{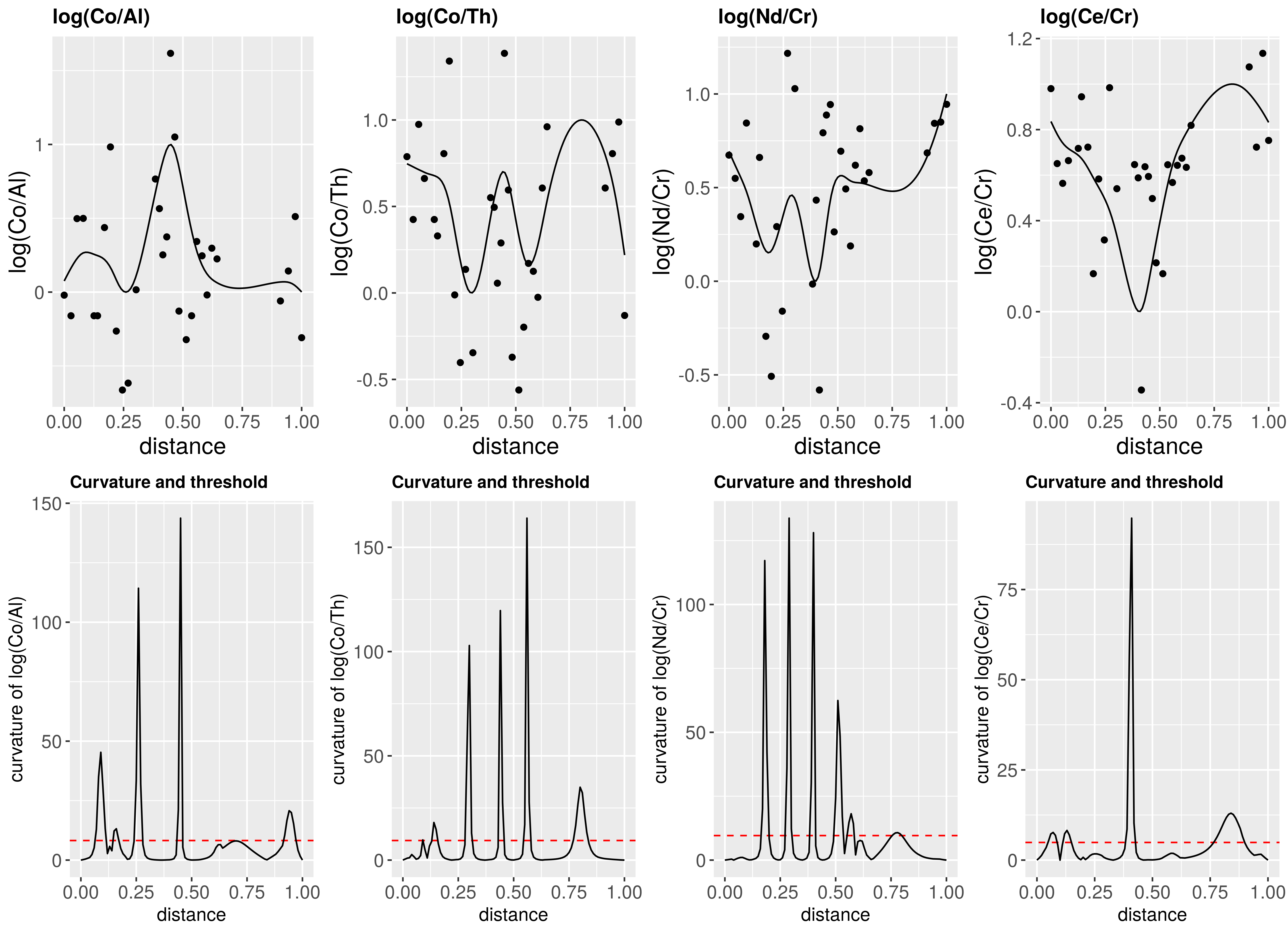}  
\caption{Upper part: four different log-ratios of GAM fits. Lower part: corresponding curvature 
together with the threshold (dashed red line).}
\label{fig:logcurv_juom}
\end{figure}

Once the GAM fits are available for all elements, their log-ratios for all different 
element pairs can be computed, together with the curvature measure.
Figure~\ref{fig:logcurv_juom} shows four examples of such log-ratios, their curvature, and the 
corresponding thresholds (dashed lines). These examples indeed reveal some of the 
known mineralizations, shown by large spatial variability which is reflected by high curvature. Note that due to the use of log-ratios, we are not necessarily interested in 
high peaks but also in low ones.

Because of their relatively high curvature values and very narrow peaks, the
pairwise log-ratios shown in Figure~\ref{fig:logcurv_juom} have high values for our $c$-value
measure defined in Equation~(\ref{measure}).
In fact, these $c$-values are assigned to the first top ranked 6 log-ratios among 351 log-ratios available in total for the particular sample material. Table~\ref{tab:rank_lr_c_juom} presents the pairwise log-ratios for the top ten curvature
measures for each sample material. Here, the measures have been scaled to the interval $[0,1]$ first
(for each sample material individually) for reasons of comparability. 
For example, for Crowberry-twig one can see that Cobalt is involved often
in the top ranked log-ratios, and thus this element seems to be a ``pathfinder'' for mineralization. Indeed, Cobalt plays an important role concerning Gold deposits. Depending on their element uptake, different plant materials can involve different elements
in the top log-ratio pairs.

\begin{table}[!htp]
\begin{center}
\caption{Top 10 ranked log-ratios and its scaled $c$-values for the plant
materials Crowberry (CRO), Bilberry (BIL), and Labrador tea (LBT).}
\label{tab:rank_lr_c_juom}
\footnotesize{
\begin{tabular}{c|cc|cc|cc|cc|cc|cc}
  & \multicolumn{2}{c|}{CRO-twig} & \multicolumn{2}{c|}{CRO-leaf} & \multicolumn{2}{c|}{BIL-twig} & \multicolumn{2}{c|}{BIL-leaf} & \multicolumn{2}{c|}{LBT-twig} & \multicolumn{2}{c}{LBT-leaf} \\
  \hline
  & pair     & $c$   & pair     & $c$    & pair     & $c$    & pair   & $c$    & pair     & $c$   & pair    & $c$  \\
  \hline
1 & Co/Al & 1 & Au/As & 1 & Se/Na & 1 & La/Th & 1 & As/Th & 1 & Se/Ag & 1 \\
2 & Co/Fe & 0.95 & As/Sc & 0.99 & Ba/Se & 0.92 & U/Fe & 0.86 & Co/As & 0.93 & Ag/Al & 0.82  \\
3 & Co/Ce & 0.84 & Bi/Pb & 0.99 & Fe/Se & 0.9 & U/S & 0.81 & As/Mo & 0.88 & Ce/Ag & 0.81 \\
4 & Co/Th & 0.78 & As/Ti & 0.95 & Se/Th & 0.83 & U/Al & 0.79 & As/Sc & 0.78 & La/Ag & 0.78 \\
5 & Ce/Cr & 0.73 & Nd/As & 0.93 & S/Se & 0.8 & U/Ni & 0.77 & As/Fe & 0.74 & Ag/Sc & 0.77 \\
6 & Nd/Cr & 0.73 & As/Cr & 0.9 & Pb/Se & 0.76 & Cu/U & 0.73 & U/As & 0.74 & Ag/Y & 0.73 \\
7 & Co/Cr & 0.72 & As/Ce & 0.9 & Se/Al & 0.76 & U/V & 0.7 & Co/Th & 0.73 & Fe/Ag & 0.73 \\
8 & La/Cr & 0.67 & As/Al & 0.88 & Se/Ti & 0.76 & U/Na & 0.63 & As/Ce & 0.7 & Nd/Ag & 0.72 \\
9 & Co/V & 0.67 & As/Na & 0.88 & Au/Se & 0.7 & U/La & 0.5 & As/La & 0.67 & Cu/Ag & 0.7 \\
10 & Co/La & 0.64 & Bi/Ni & 0.88 & Ce/Se & 0.7 & U/Th & 0.5 & As/Y & 0.66 & Ag/Ti & 0.67 \\
\end{tabular}
}
\end{center}
\end{table}


Similarly, the interpretation for individual combinations of plants can be provided.
This information is stored in Table \ref{tab:rank_lr_el_juom}, where the column ``element'' 
provides the first top 10 elements which appear most often in the best ranked log-ratios for
the particular plant media. For example, for plant species Crowberry and plant
tissue twig, the elements Co, As,
Cr, Bi, Nd (in this order) most frequently appear among the best ranked log-ratios. 
These elements can thus be considered as pathfinder elements.

\begin{table}[ht]
\centering
\caption{Top 10 ranked log-ratios and its elements for each material.}   
\label{tab:rank_lr_el_juom}
\footnotesize{
\begin{tabular}{c|c|c|c|c|c|c}
  & \multicolumn{1}{c}{CRO-twig} & \multicolumn{1}{c}{CRO-leaf} & \multicolumn{1}{c}{BIL-twig} & \multicolumn{1}{c}{BIL-leaf} & \multicolumn{1}{c}{LBT-twig} & \multicolumn{1}{c}{LBT-leaf} \\
  \hline
  & element   &  element    &  element    &  element       & element      & element    \\
  \hline
1 & Co &  As &  Se &   U &  As &  Ag \\
2 & As &  Bi &  U &   Bi &  Co &  Y   \\
3 & Cr &  Co &  W &  La &  U &  Nd    \\
4 & Bi &  Sc &  Tl &  Th &  Fe &  Ce \\
5 & Nd &  Cr & Ag &   Cr &  W &  Au   \\
6 & Ce &  Fe &  Co &  Fe & Th & Al   \\
7 & Fe &  Al &  Fe & S &  Se &  Co   \\
8 & La &  Y &  Na &   Ce &  Al &  Fe  \\
9 & Al &  La &  Bi &  Y &  Mo &  La   \\
10 & Y &  Ti &  Ba &  Cu & Ag &  U   \\
\end{tabular}
}
\end{table}

\begin{figure}[htp]
\centering
\includegraphics[width=1\textwidth]{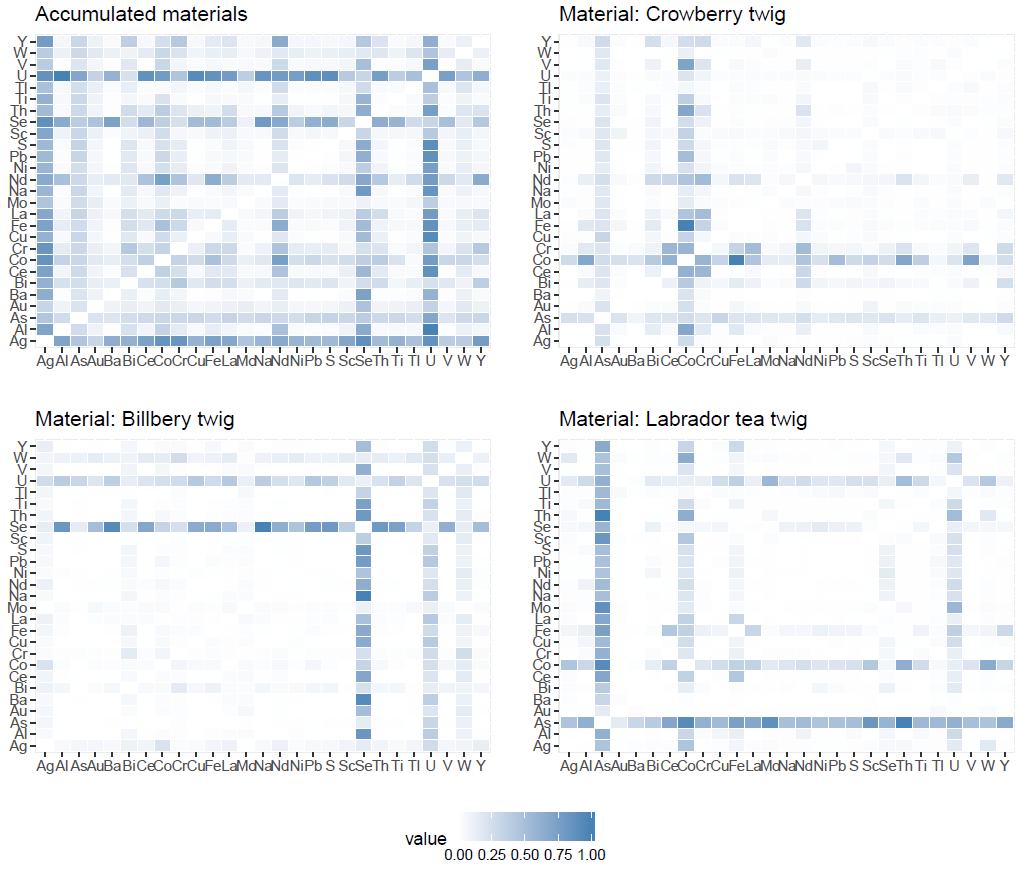}
\caption{Heatmaps of the $c$-values per element for all possible log-ratios 
of the tissue twig for all plant species, and the accumulated values of all materials 
(upper left).}
\label{fig:heat_twig_juom}
\end{figure}

The information contained in the ranked lists can also be visually summarized in heatmaps.
The scaled $c$-value measures need to be mapped to colors, where in the following representation
0 was mapped to white, and 1 to dark blue, with a continuous
spectrum between these extremes.
Figure~\ref{fig:heat_twig_juom} shows the resulting heatmaps for the sample media twigs of 
the different plant species, as well as a heatmap for the accumulated values 
of all sample materials (upper left). The heatmaps represent the different elements in the 
rows and columns, with symmetry around the diagonal. Each cell in the heatmap refers to
the scaled $c$-values of the corresponding pairwise log-ratio. For instance, the upper left
plot for accumulated panel shows that Silver (Ag) is involved in many log-ratios with high values 
of the $c$-values.
Also Arsenic (As), Bismuth (Bi), Cobalt (Co), Selenium (Se) and Uranium (U) are present
in many important log-ratios. The interpretation of those mentioned elements can be partially seen in \cite{middleton2018biogeochemical}. Ag, Bi and Se are elements verified by lithogeochemistry and also elements exhibiting anomalous spatial patterns over the mineralization. Cobalt creates together with Gold the underlying hydrothermal deposit, where Uranium is one of the elements showing spatial multi-elemental anomaly patterns for Au-Co deposits. Arsenic is one of the interesting pathfinder elements from the perspective of geochemical exploration.
The heatmaps for the individual plant materials provide different information, because 
it depends very much on the plant materials which elements are enriched by a potential
mineralization. 
The heatmap of the sample medium Crowberry-twig (upper right) shows a couple of highly 
ranked log-ratios indicated by dark blue color, where Cobalt seems to be involved in 
several log-ratios, for instance log(Co/Al), log(Co/Fe), log(Co/Ce), log(Co/Th), etc. One of the conclusions in \cite{middleton2018biogeochemical} is that Crowberry twigs were the most efficient plant tissues for revealing the location of the mineralized lodes determined as background for Au-Co deposits. Another investigated plant was Bilberry twig (lower left heatmap) clearly showing two elements with high $c$-values, i.e. Selenium and Uranium. 
In fact, it turned out that Bilberry twig was rather a poor quality indicator for many elements, forming spatial anomaly patterns for only a few elements. The last heatmap (lower right) for Labrador tea and its twig clearly shows Arsenic as the mostly involved in highly ranked log-ratios.
Arsenic belongs to the group of so called pathfinder elements; another important element seems to be Cobalt and Uranium forming the group of ore elements.

\subsection{Gj{\o}vik Data}

As a second application we use the Gj{\o}vik data set, which
originates from a project of the Geological Survey of Norway (NGU) in a 100 km transect in 
Gj{\o}vik, Norway \citep{reimann2018response}. 
In total, 15 different sample materials have been investigated, soil as well as plants,
and approximately 40 samples are available for each subdata set. They have been sampled more or less on a
linear transect, and for our method we first derived the distances between the samples
by projection onto a line. We selected 30 chemical elements with reasonable data quality.
The GAM fits have been computed for each element and each sample material, followed by
computing the log-ratios and the curvature measure.

\begin{figure}[ht]
\centering
\includegraphics[width=1\textwidth]{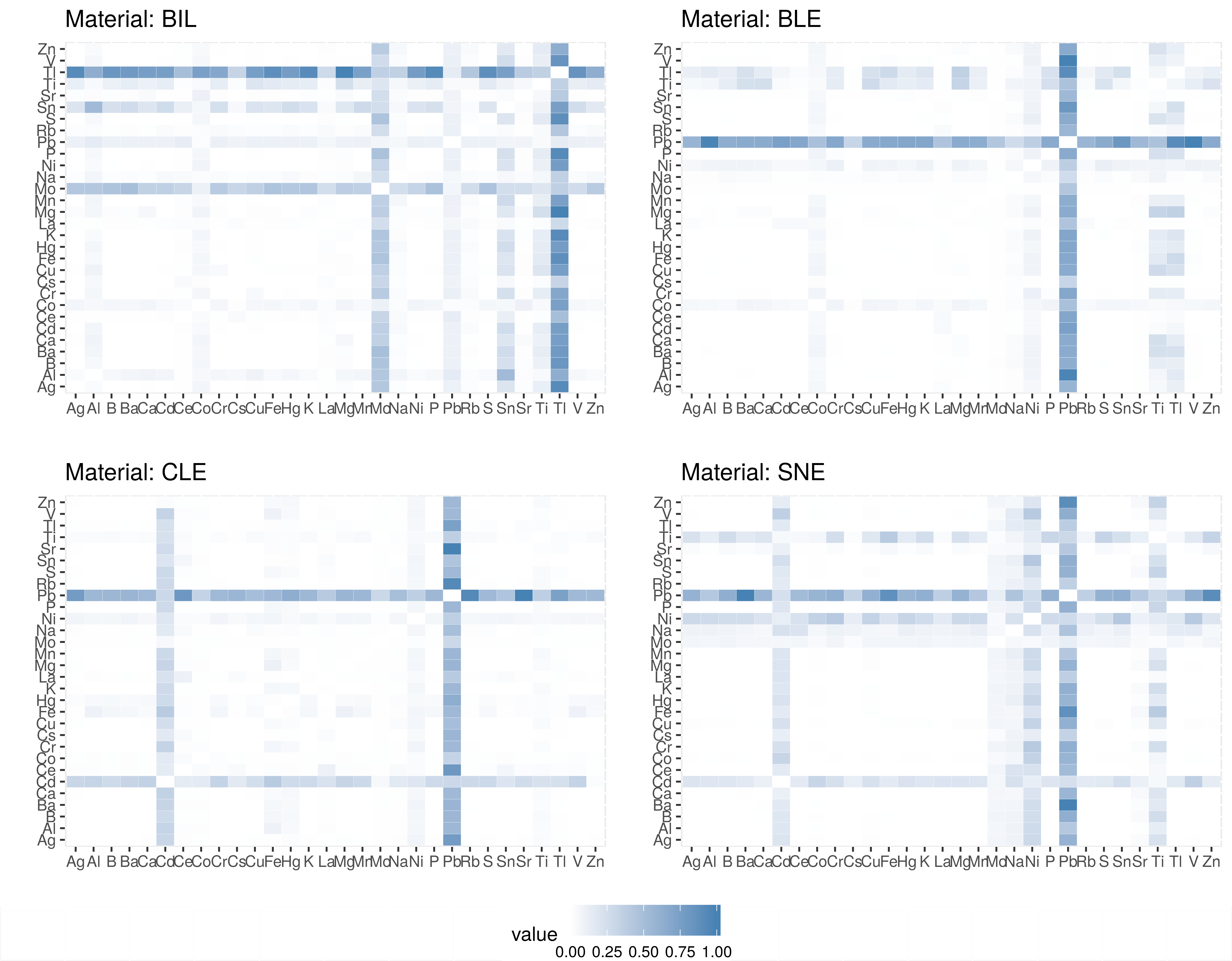}
\caption{Heatmaps of the $c$-values for all possible log-ratios of four media -- BIL, BLE, CLE, SNE.}
\label{fig:heat_gjovik}
\end{figure}

Figure~\ref{fig:heat_gjovik} shows the resulting heatmaps for four selected sample materials,
Birch leaves (BIL), Blueberry leaves (BLE), Cowberry leaves (CLE), and Spruce needles (SNE).
Three of these plots show that Lead (Pb) seems to be a pathfinder element, but also 
Tl (Thallium), Mo (Molybdenum), Sn (Tin), and Ti (Titanium) result in log-ratios with 
high $c$-values.
In fact, the Gj{\o}vik data set has been investigated because there are known mineralizations
of Lead (Pb) and Molybdenum (Mo).

\begin{figure}[ht]
\centering
\includegraphics[width=1\textwidth]{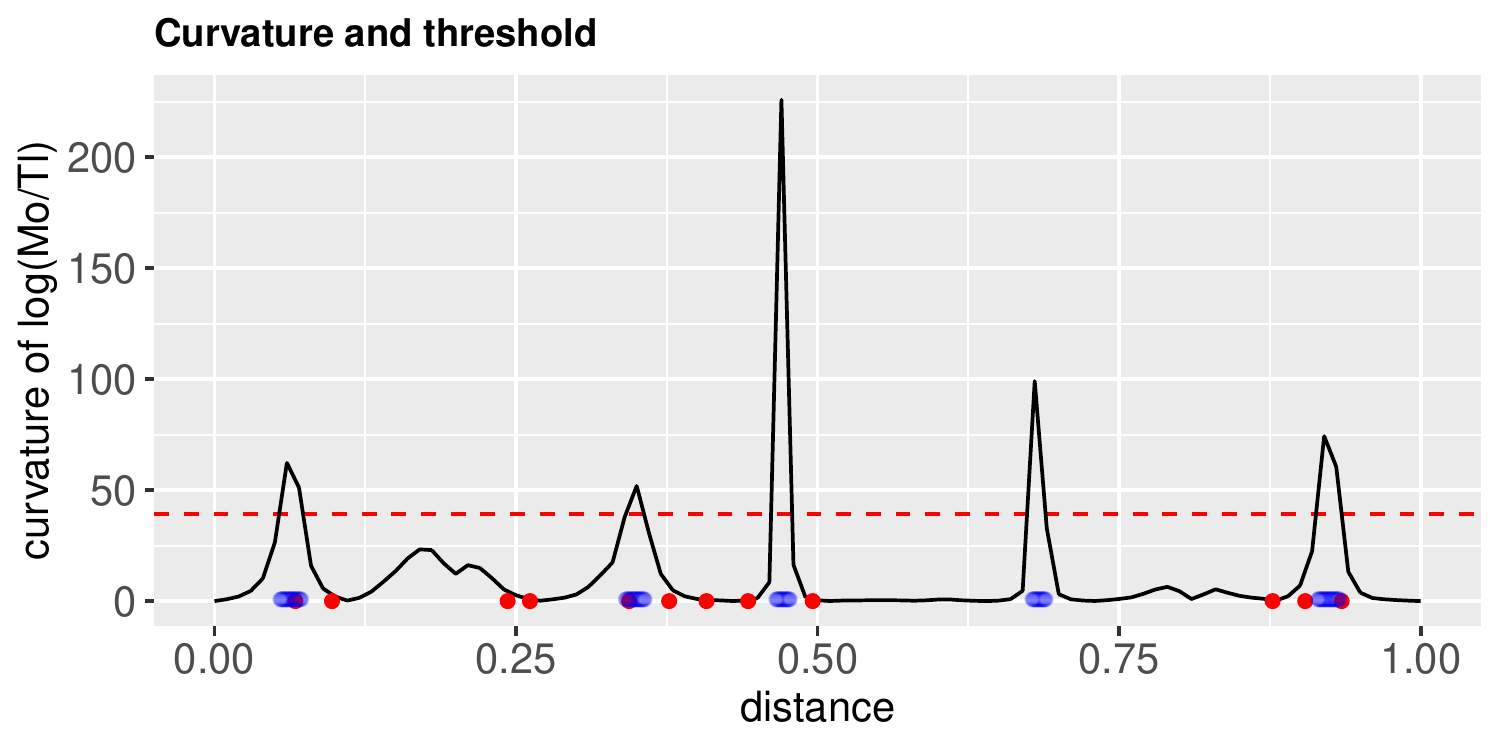}
\caption{Curvature of log(Mo/Tl) in sample material Birch leaves, 
where known mineralization area (red points) and mineralized points identified by the method (blue points) are displayed.}
\label{fig:gjovik_yes_no}
\end{figure}

Figure~\ref{fig:gjovik_yes_no} focuses on the two elements Mo and Tl for the 
sample material Birch leaves. Both elements may be relevant for identifying mineralization.
The solid line in the plot is the 
log-ratio of the GAM fits for these two elements, and the dashed line corresponds to the threshold used 
inside the algorithm to compute the curvature measure. 
The blue points indicate the predicted areas of mineralization, while the red points indicate the 
known mineralizations for Lead and Molybdenum. 
There is a strong overlap of the known and predicted areas, and in addition to that there might
be new predicted areas worthwhile to be explored.

\begin{figure}[H]
\centering
\includegraphics[width=0.75\textwidth, page = 2]{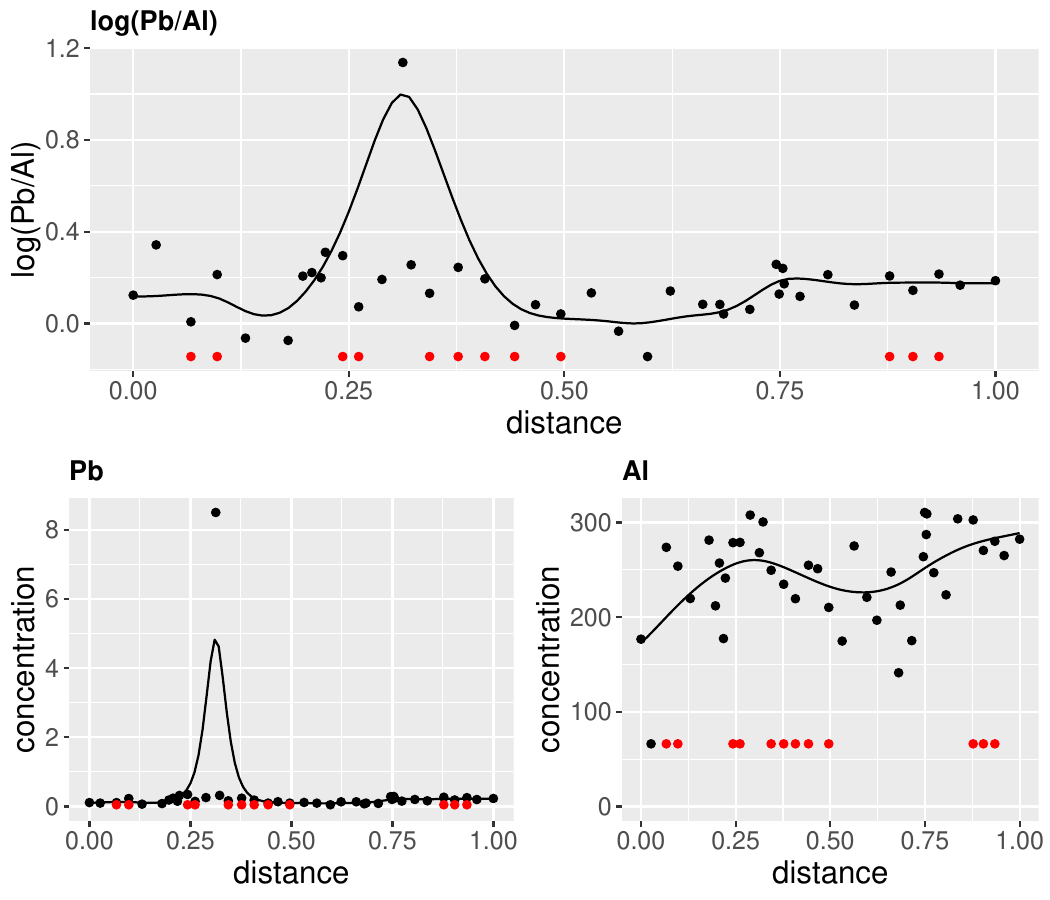}
\caption{Upper part: Log-ratio of Lead and Aluminium constructed by using GAM fits of its individual elements -- displayed on lower part of plot. Sample material is BLE. The red points indicate areas of known mineralization.}
\label{fig:gjovik_minpoint}
\end{figure}

Another example of an interesting log-ratio is displayed in Figure~\ref{fig:gjovik_minpoint}.
This log-ratio of Lead (Pb) versus Aluminium (Al) is ranked as the second most important 
log-ratio (according to our $c$-value) in sample material BLE.
In this figure we also show the original measured concentrations for Pb and Al, their 
GAM fits, and the locations of the known Lead anomalies (red points). The log-ratio 
of the GAM fits clearly indicates the area around the known mineralization. The second known
Pb mineralization around distance 0.9 is not indicated. This is because there is no increased
measured Pb value around this distance, or the sampling survey has missed an appropriate
measurement.

\begin{figure}[H]
\centering
\includegraphics[width=0.78\textwidth
]{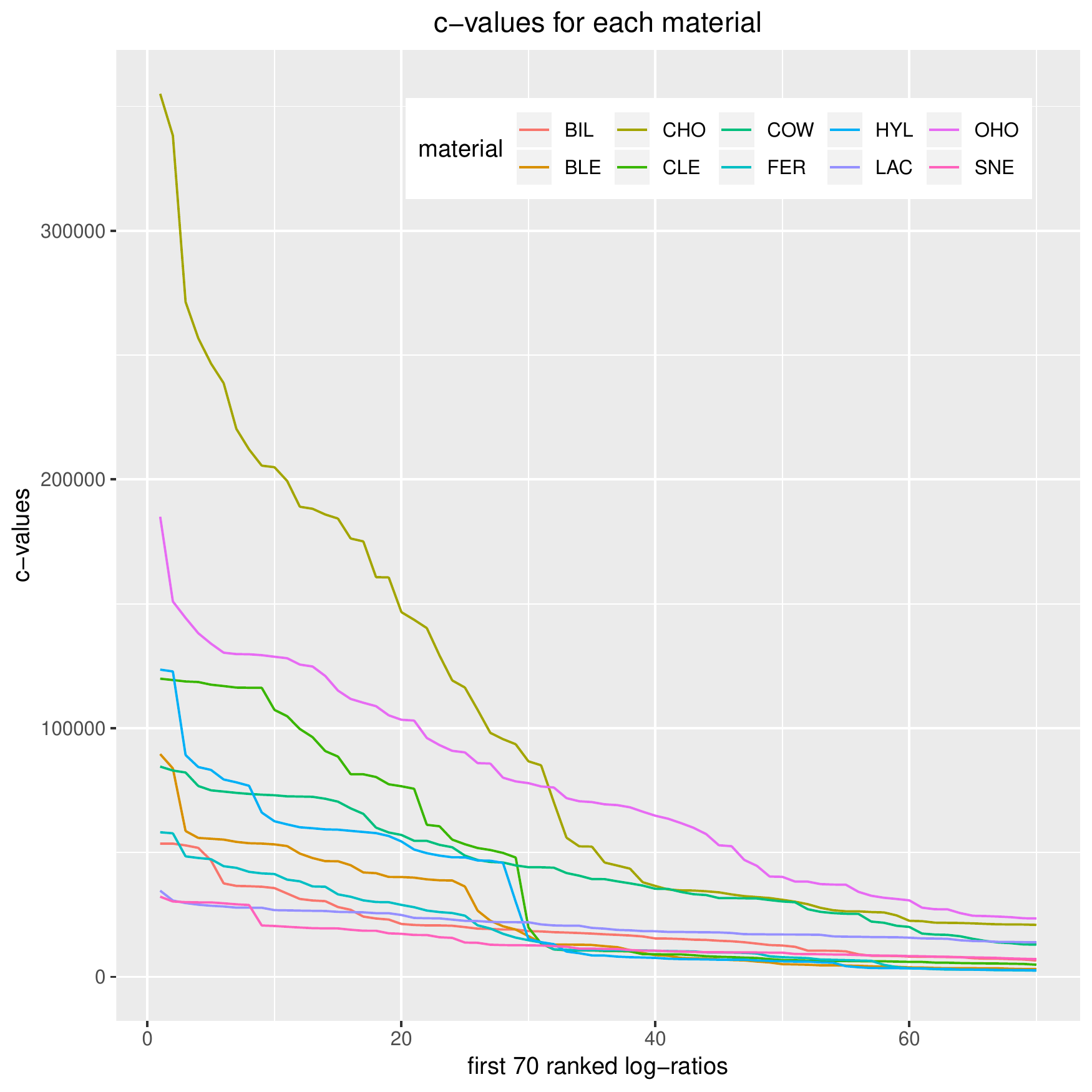}
\caption{Top-ranked 70 (unscaled) $c$-values for each sample material.
The horizontal axis represents the rank.}
\label{fig:lines_gjovik}
\end{figure}

In order to stress the importance of the individual sample materials, Figure~\ref{fig:lines_gjovik} displays the 70 top ranked $c$-values (unscaled) of these materials. 
Every line in the plot corresponds to a particular material, and the top-ranked $c$-values 
are connected by the line. The highest $c$-values are obtained for the O-horizon (OHO)
samples. A quick decline of the curve means that the $c$-values of lower-ranked log-ratios 
are clearly smaller than for the top-ranked ones.
One still needs to be careful with the interpretation, because high $c$-values can be obtained
by few wide peaks in the log-ratios of the GAM fits, and not necessarily by several 
sharp peaks. In practice it will be definitely worthwhile to inspect the top-ranked log-ratios
for several sample media, based on the information of Figure~\ref{fig:lines_gjovik}.

\newpage
\section{Discussion and conclusions}

The identification of mineralization is usually based on a pre-study in the prospective
area, where only few sample locations are considered. The sample locations are supposed
to cross the potential mineralized zones, and thus the samples are frequently arranged
on a linear transect. This is the setting which we considered in this paper.

Due to the compositional nature of geochemical data, log-ratios of the element concentrations
are considered as informative. A further important property is the scale-invariance of
log-ratios, which is very important when comparing log-ratios of different elements.
However, analyzing log-ratios of the measurements of only few sample 
sample points may lead to a lot of uncertainty and instability. For this reason, the 
concentration data have been smoothed first using Generalized Additive Models (GAMs).
The advantage of GAMs is that the smoothness can be tuned with a parameter, and the tuning
parameter is selected using the underlying data (with cross-validation). Thus, the smoothing
is adapted to the data, and once the smoothed signal is available, an arbitrary number of 
``concentration'' values can be generated. Taking log-ratios of such generated values 
allows to compute the curvature, which involves the first and the second derivative, and these 
can be numerically obtained. Finally, a measure of ``overall'' curvature, which we called 
$c$-value, can be obtained. The $c$-value is not depending on the measurement units,
and it can thus be compared for different element combinations, and even accross different
sample media. Moreover, due to the symmetry of log-ratios, an exchange of nominator and 
denominator would result in exactly the same $c$-value, which reduces the number of potential
element combinations for the algorithm significantly.

In the experimental part we have demonstrated using two geochemical mineral exploration
studies, that this methodology is indeed promising to identify pathfinder elements
for mineralization. Those elements that appear in the top-ranked log-ratios (ranking 
according to the $c$-value) are considered as most informative. In addition, the inspection
of the top-ranked log-ratios gives an indication of the location of the mineralization, 
and even of the extent of the mineralized areas. Furthermore, it is informative to 
inspect the magnitude of the top-ranked $c$-values for the different sample media in order 
to get an idea about their importance for the task of mineral exploration.

It is important to mention that the algorithm is unsupervised. This means that prior knowledge
on the mineralized locations is not necessary. In our studies we only used this prior
knowledge for the verification of the results.

In our future work we will extend this approach to the two-dimensional setting, i.e.~where the
samples are not taken along a linear transect but at locations in a two-dimensional
(irregular) grid. Although computationally more challenging, GAM fits can be extended to the two-dimensional
case, but curvature also needs to be extended to this setting, together with an appropriate 
measure of overall curvature.

\section*{Acknowledgments}

The authors acknowledge support from the UpDeep project 
(Upscaling deep buried geochemical exploration techniques into European 
business, 2017-2020), funded by the European Information and Technology Raw Materials. We thank GTK personel Janne Kivilompolo and Kari Kivilompolo for mountain birch sampling and Jukka Konnunaho for organizing the funding and project management of Mineral potential of northern Finland. 

\vspace{1cm}

\newpage
\bibliographystyle{plainnat}
\bibliography{ref}

\end{document}